\newcommand{\aj}{AJ}
\newcommand{\apj}{ApJ}
\newcommand{\apjl}{ApJL}
\newcommand{\apjs}{ApJS}
\newcommand{\aap}{A\&A}

\newcommand{\mnras}{MNRAS}
\newcommand{\araa}{ARA\&A}

\newcommand{\hdf}{HDF--N}

\newcommand{\hst}{\textit{HST}}
\newcommand{\spitzer}{\textit{Spitzer}}

\newcommand{\acsb}{\hbox{$B_{435}$}}
\newcommand{\acsv}{\hbox{$V_{606}$}}
\newcommand{\acsi}{\hbox{$i_{775}$}}
\newcommand{\acsz}{\hbox{$z_{850}$}}

\newcommand{\ks}{\hbox{$K_s$}}

\newcommand{\lsim}{\lesssim}
\newcommand{\gsim}{\gtrsim}

\newcommand{\mcal}{\hbox{$\mathcal{M}$}}

\newcommand{\etal}{et al.}
\newcommand{\eg}{e.g.}
\newcommand{\ie}{i.e.}

\newcommand{\msol}{\hbox{$\mathcal{M}_\odot$}}

\newcommand{\lsol}{\hbox{$L_\odot$}}

\newcommand{\jmk}{\hbox{$J - K_s$}}

\newcommand{\ujy}{\hbox{$\mu$Jy}}
\newcommand{\lir}{\hbox{$L_{\mathrm{IR}}$}}

\newcommand{\micron}{\hbox{$\mu$m}}
\newcommand{\arcmin}{\hbox{$^\prime$}}

\documentclass{elsart}
\usepackage{natbib}
\usepackage{graphicx}
\usepackage{amssymb}

\begin{document}

\begin{frontmatter}



\title{Spitzer Observations of Red Galaxies: Implication for
High--Redshift Star Formation}

\vspace{-20pt}
\author[label1]{Casey Papovich}
\ead{papovich@as.arizona.edu}
\author{for the GOODS and MIPS GTO teams}
\vspace{-5pt}
\address[label1]{Spitzer Fellow, Steward Observatory, 933 N.\ Cherry Avenue, Tucson, AZ 87521}
\vspace{-10pt}
\begin{abstract}
% Text of abstract
My colleagues and I identified distant red galaxies (DRGs) with
\jmk$>$2.3 in the southern Great Observatories Origins Deep
Surveys (GOODS--S) field. These galaxies reside at $z$$\sim$1--3.5,
($\langle z\rangle$$\simeq$2.2) and based on their ACS
(0.4--1~\micron), ISAAC (1--2.2~\micron), and IRAC (3--8~\micron)
photometry, they typically have inferred stellar masses
$\mcal$$\gsim$$10^{11}$~\msol.  Interestingly, more than 50\% of these
objects have 24~\micron\ flux densities $\geq$50~\ujy.   Attributing
the IR emission to star--formation implies star--formation rates
(SFRs) of $\simeq$100--1000~\msol\ yr$^{-1}$.   As a result, galaxies
with $\mcal$$\geq$ $10^{11}$~\msol\ have specific SFRs equal to or
exceeding the global value at $z$$\sim$1.5--3.  In contrast, galaxies
with $\mcal$$\geq$$10^{11}$~\msol\ at $z$$\sim$0.3--0.75 have specific
SFRs less than the global average, and more than an order of magnitude
lower than that for massive DRGs at $z$$\sim$1.5--3.  Thus,  the bulk
of star formation in massive galaxies is largely complete by
$z$$\sim$1.5.    The red colors and large inferred stellar masses in
the DRGs suggest that much of the star formation in these galaxies
occurred at redshifts $z$$\gsim$5--6.   Using model star--formation
histories that match the DRG colors and stellar masses at
$z$$\sim$2--3, and measurements of the UV luminosity density at
$z$$\gsim$5--6, we consider what constraints exist on the stellar
initial mass function in the progenitors of the massive DRGs at
$z$$\sim$2--3.
\end{abstract}
\vspace{-6pt}

\end{frontmatter}
\vspace{-36pt}

% main text
\section{Introduction}\label{section:intro}\vspace{-24pt}
Although as much as $\sim$50\% of the stellar mass in galaxies today
may have formed during the short time between $z$$\sim$3 and 1
\citep[\eg][]{dic03,rud03}, it is still unclear
where these stars formed.  E.g., one hypothesis is that galaxies
``downsize'', with massive galaxies forming most of their stars in
their current configuration at high--$z$, and lower mass galaxies
continuing to form stars at lower--$z$
\citep[\eg,][]{cow99,bauer05,delucia05,jun05}.   Alternatively, stars
may form predominantly in low--mass galaxies at high--$z$, which
then merge hierarchically over time, slowly assembling into large,
massive galaxies at more recent times \citep[\eg,][]{bau98,kau98,cim02b}.

Massive galaxies with stellar populations older than a few megayears
have recently been identified and studied with \spitzer\ as high as
$z$$\sim$6--7 \citep{eyl05,yan05,mob05}.  One question from this is,
what are the descendants of these $z$$\sim$6 massive galaxies?  And,
perhaps more importantly (especially in the context of this meeting),
what kind of stellar populations are forming in these galaxies (i.e.,
are their enough massive stars to produce the Lyman--continuum photons
necessary to reionize the Universe)?  It appears that there are too
few luminous quasars at $z$$\gsim$5--6 to provide enough global
radiation for reionization \citep{fan02}.   So, the burden now rests
on early stellar generations to provide enough UV photons for
reionization.  It may be that there are enough small galaxies at
$z$$\gsim$6 with high UV escape fractions for this task
\citep{yan03}, or maybe the old stellar populations in massive
$z$$\sim$6 galaxies supplied these photons \citep{pan05}.  There are
as--of--yet few constraints on the stellar populations in these
galaxies.

In these proceedings, I discuss recent \spitzer\ observations at
(3--24~\micron) of massive galaxies at $z$$\sim$1.5--3 in the
southern Great Observatories Origins Deep (GOODS--S) Field.   The
\spitzer/MIPS 24~\micron\ observations provide new constraints on the SFRs
in massive galaxies at these epochs.  In addition, the \spitzer/IRAC
observations at 3.6--8~\micron\ constrain the \textit{total}  mass in
old stars in galaxies at these redshifts.  As I argue below, the
constraints on the mass in stars from ancient stellar populations now
provide crude limits on the form of the initial mass function (IMF)
in galaxies near the epoch of reionization.

\vspace{-24pt}
\section{\spitzer\ Observations of the Massive Galaxies in
GOODS--S}\label{section:data}
\vspace{-24pt}

GOODS is a multiwavelength survey of two 10\arcmin$\times$15\arcmin\
fields, one in the northern \textit{Hubble} Deep Field, and one in the
southern \textit{Chanrda} Deep Field.  The GOODS datasets include
(along with other things)
\hst/ACS and VLT/ISAAC imaging \citep{gia04a}, and recent \spitzer\
imaging (M.\ Dickinson et al., in prep.).  We also use data from
\spitzer/MIPS 24~\micron\ in the GOODS--S field from time allocated to
the MIPS GTOs \citep{pap04b}.    

From the GOODS--S datasets,  my colleagues and I identified distant
red galaxies (DRGs) with \jmk$>$2.3~mag. \citet{fra03} used this color
criterion to identify galaxies at $z$$\sim$2--3.5 that in principle
have a strong Balmer/4000~\AA\ break between the $J$-- and \ks--bands.
This color selection should be sensitive to any galaxy dominated by a
passively evolving stellar population older than $\sim$250~Myr
at these redshifts \citep{fra03}.  For the GOODS--S ISAAC images, the
\jmk$>$2.3~mag selection is approximately complete to stellar mass
$\mcal$$\geq$$10^{11}$~\msol\ for passively evolving galaxies.  In the
GOODS--S, we find 153 DRGs to $\ks$$\leq$23.2~mag, spanning
0.8$\leq$$z$$\leq$3.7, with $\langle z\rangle $$\simeq$2.2.  Roughly
15\% of these have X--ray detections in 1~Msec \textit{Chandra} data
\citep{ale03}, suggesting that many of these galaxies harbor
strong AGN (see Papovich et al., 2005; L.~Moustakas et al., in prep.).

Approximately 50\% of the DRGs have \spitzer\ detections with
$f_\nu(24\micron)$ $\geq$50~\ujy.  The 24~\micron\ emission at these
redshifts probes the mid--IR ($\sim$5--10~\micron), which broadly
correlates with the total IR, $\lir$$\equiv$$L(8-1000\micron)$
\citep[\eg,][]{cha01}.  We convert the observed 24~\micron\ flux
densities to a total IR luminosity using the \citet{dal02} IR template
spectral energy distributions and the measured redshifts.  Some
scatter is inherent in this converstion, and is typically a factor of
2--3 in \lir\ \citep[\eg,][]{cha03}.   We therefore adopt a 0.5~dex
uncertainty on \lir, which also includes an uncertainty arising from
redshift errors \citep[see][]{pap05}.   For the DRGs at
$z$$\sim$1.5--3, the 24~\micron\ flux densities yield
$\lir$$\approx$$10^{11.5-13}$~\lsol, which if attributed to
star--formation corresponds to SFRs of $\approx$100--1000~\msol\
yr$^{-1}$ \citep{ken98,bel03}.  Thus, a substantial fraction of
massive galaxies at these redshifts are involved in intense
starbursts.  A similar conclusion is reached for $BzK$--selected
galaxies at $z$$\sim$1.9 in the northern GOODS field
\citep{dad05}.  

\vspace{-24pt}
\section{Stellar Populations and SFRs in High--$z$ Massive
Galaxies}\label{section:modeling}
\vspace{-24pt}

\begin{figure}[t] 
\begin{minipage}[t]{70mm} 
\begin{flushleft}
%\begin{center}
\includegraphics*[width=75mm]{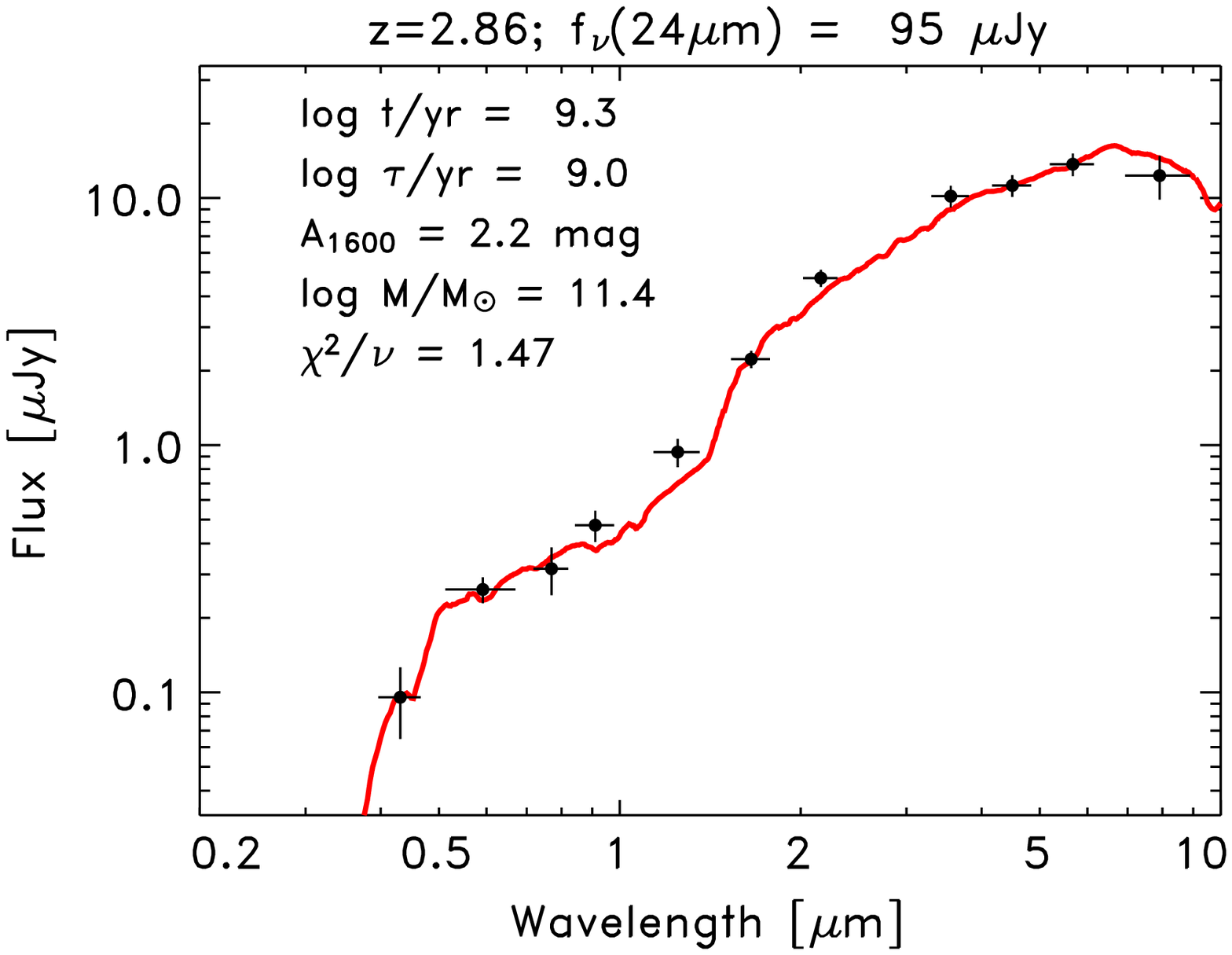}
%\end{center}
\end{flushleft}
\end{minipage} 
%\hspace{\fill} 
\begin{minipage}[t]{70mm} 
\begin{flushleft}
%\begin{center}
\includegraphics*[width=75mm]{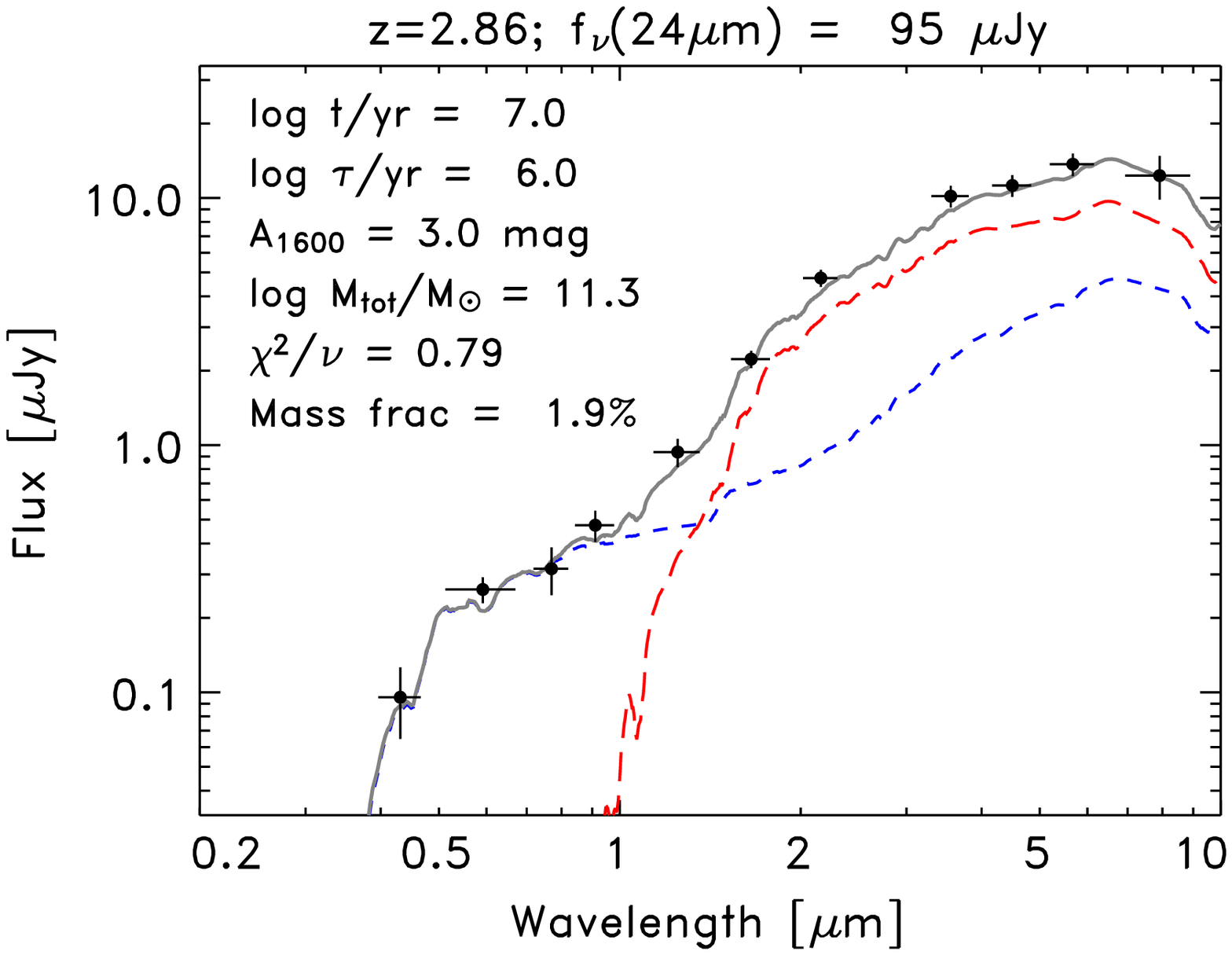}
%\end{center}
\end{flushleft}
\end{minipage} 
\caption{Illustration of
fitting SED models to one of the GOODS--S DRGs
which has an indication of both young and old stellar populations.
\textit{LEFT:} the best--fit model with the star--formation history
parameterized as a monotonic, decaying exponential with best--fit
$e$--folding time $\tau$=$10^9$~yr.  \textit{RIGHT:} the best--fit
model where the star formation history consists of two components, a
single burst at $z$$=$$\infty$, representing previously formed stellar
populations, summed with a monotonic, decaying exponential with
best--fit $e$--folding time $\tau$=$10^6$~yr, representing any ongoing
star formation.   The dashed curves in the right panel show the
contribution from each of the two model components.   The inset gives
the best--fit model parameters.   The data points show the ACS
\acsb\acsv\acsi\acsz, ISAAC $JH$\ks, and IRAC 3.6--8.0~\micron\
photometry and errors.}\label{fig:example_spec}
\end{figure}

We modeled the DRG stellar populations by comparing their ACS, ISAAC,
and IRAC  photometry to a suite of stellar--population synthesis
models (Bru\-zu\-al \& Char\-lot, 2003),  varying the age,
star--formation history, and dust content.  We use the model
stellar-mass--to--light ratios to estimate the galaxies' stellar mass.
We first allow for star--formation histories with the SFR
parameterized as a decaying exponential with an $e$--folding time,
$\tau$, ranging from short $\tau$'s (burst of star--formation) to long
$\tau$'s (constant star--formation).  We also use models with a
two--component star--formation history characterized by a passively
evolving stellar population formed in a ``burst'' at
$z_\mathrm{form}$=$\infty$, summed with the
exponentially--decaying--SFR model above. The two--component models
check the effects of discrete bursts. Figure~\ref{fig:example_spec}
shows examples of fitting these models to a DRG with an apparent
4000~\AA/Balmer break between the $J$-- and \ks--bands.  The two
panels in the figure show the best--fitting models for the one-- and
two--component models described above.   Although the modeling loosely
constrains the ages, dust content, and star--formation histories of
the DRGs, it provides relatively robust estimates of the galaxies'
stellar masses \citep[see also F\"orster--Schreiber et al.,
2004]{pap05}.   In the DRG in figure~\ref{fig:example_spec}, even
though the star--formation histories are quite different, the derived
stellar mass is nearly identical.   Typical uncertainties for the
stellar masses for the full DRG sample are 0.1--0.3~dex.

\begin{figure*}[t]
\begin{center}
\includegraphics*[width=75mm]{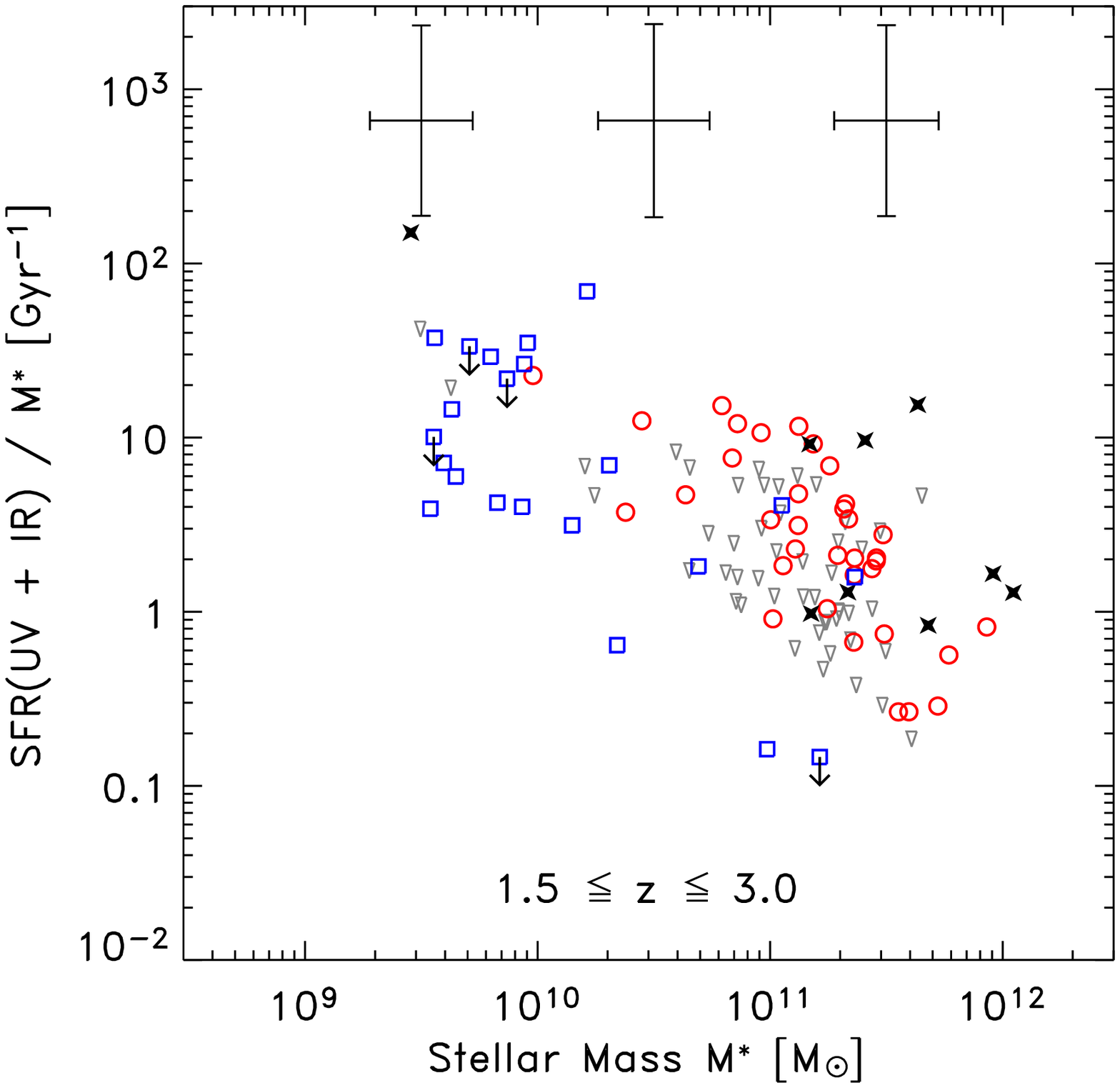}
\end{center}
\vspace{-24pt}
\begin{minipage}[t]{70mm}
\begin{flushleft}
\includegraphics*[width=75mm]{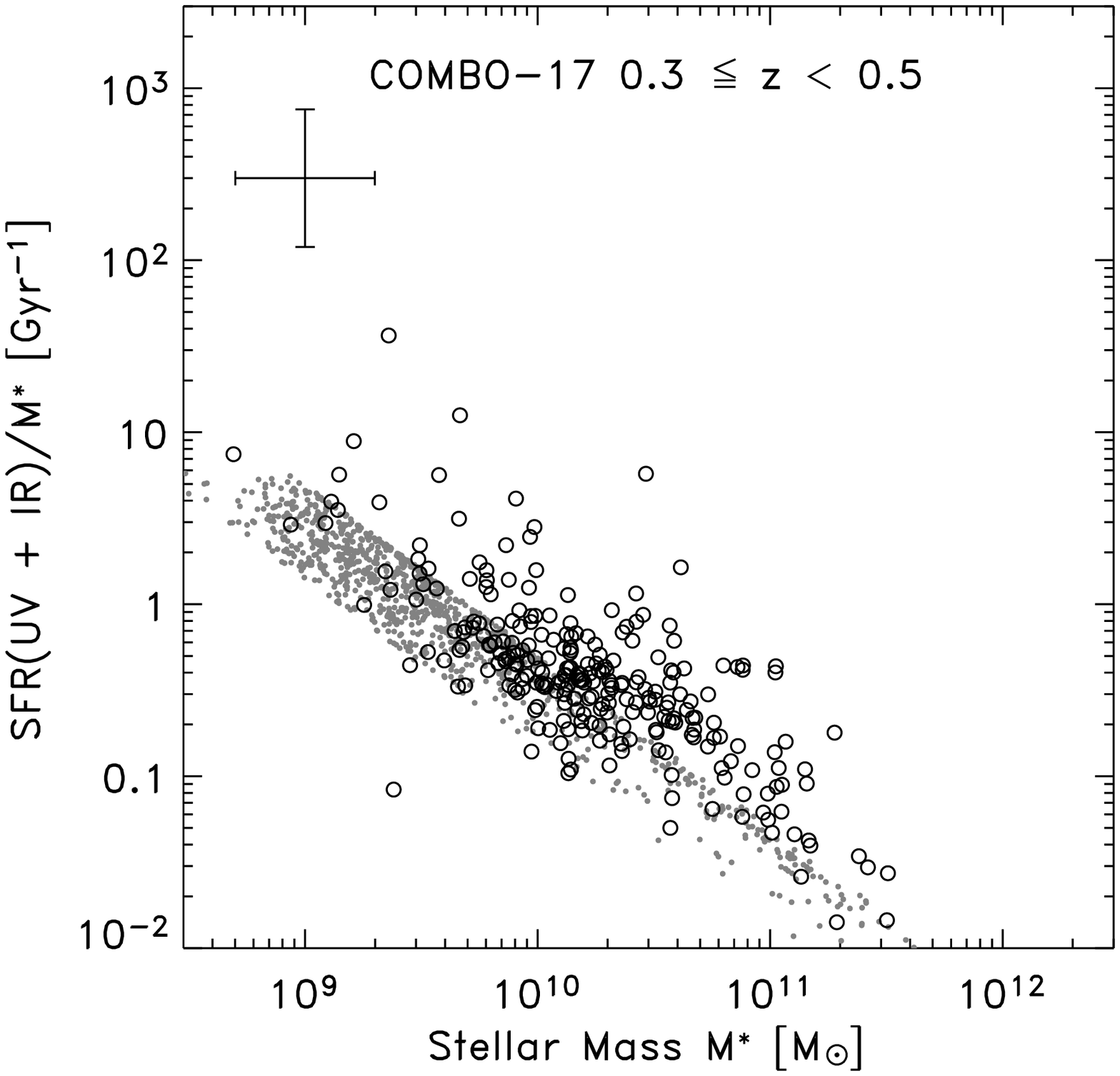}
\end{flushleft}
\end{minipage}
\hfill
\begin{minipage}[t]{70mm}
\begin{flushleft}
\includegraphics*[width=75mm]{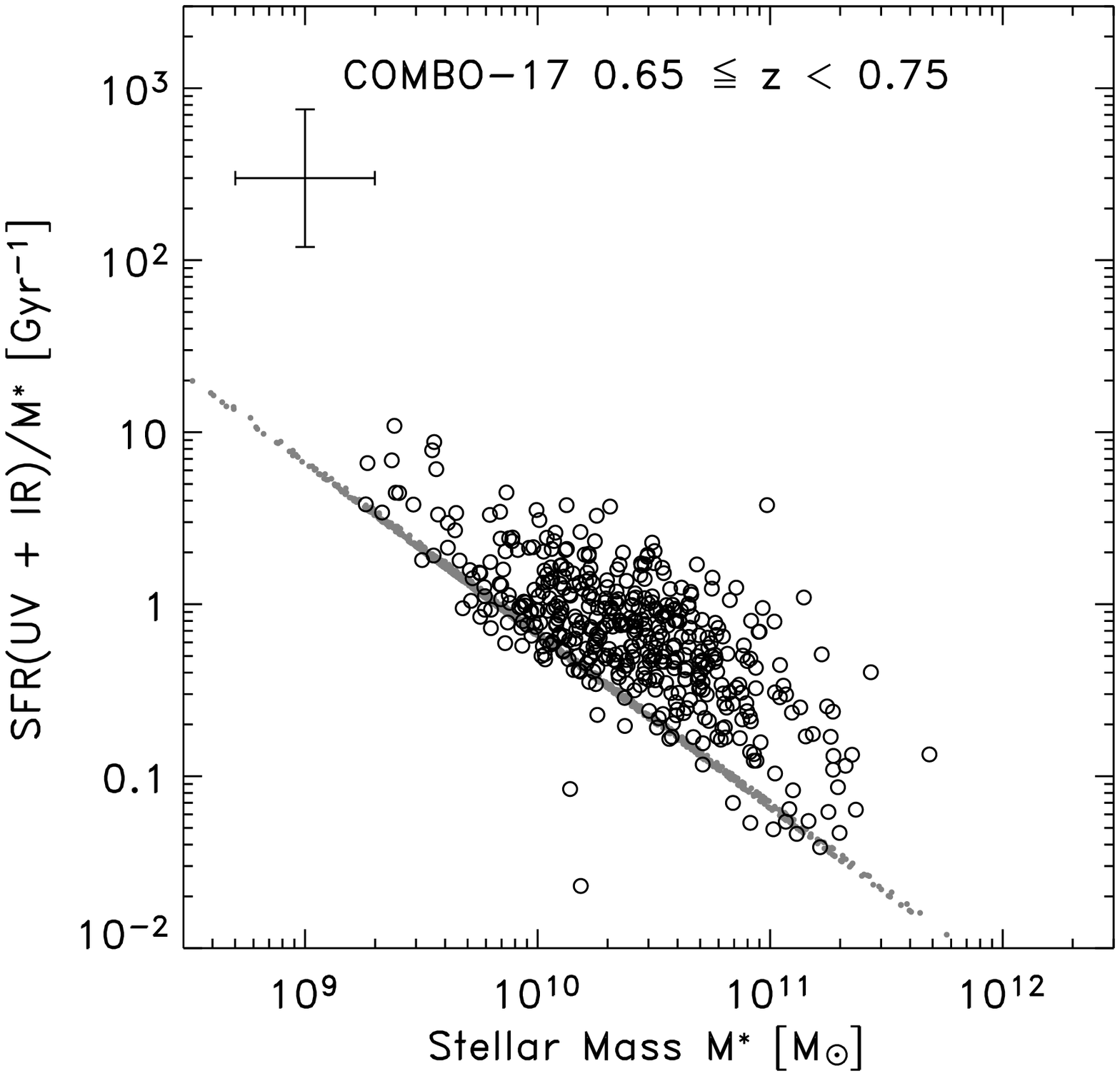}
\end{flushleft}
\end{minipage}
\caption{Galaxy stellar mass versus specific SFR.  The top
panel shows the DRGs (red circles, grey triangles
denote upper limits, black stars indicate X--ray DRGs) and galaxies
from the \hdf\ (blue squares; see Papovich et al., 2005),  restricted
to 1.5$\leq$$z$$\leq$3.0. The bottom panels show COMBO--17 galaxy
samples at lower--redshift (as labeled). Open circles show
24~\micron--detected galaxies; small filled circles show upper limits.
Mean uncertainties are indicated as a function of stellar
mass.\label{fig:specsfrmass} }
\end{figure*}

Figure~\ref{fig:specsfrmass} shows the specific SFRs ($\Psi/\mcal$)
derived from the masses and SFRs the DRGs, where the SFRs are
calculated from the summed UV and \spitzer\ IR emission \citep{bel05}.
The figure also shows the specific SFRs for galaxies at lower redshift
in the COMBO--17 survey \citep{wol03}, which overlaps with the GTO
\spitzer\ imaging at 24~\micron.   The SFRs for the COMBO--17 galaxies
are calculated using the MIPS 24~\micron\ imaging and rest--frame UV
emission in an analogous manner as for the DRGs.   Masses for
COMBO--17 galaxies were estimated from their rest--frame $M(V)$ and
$U-V$ colors, and have a typical uncertainty of 0.3~dex (see Bell et
al., 2004, 2005).

\begin{figure*}[t]
\begin{center}\includegraphics*[scale=0.5]{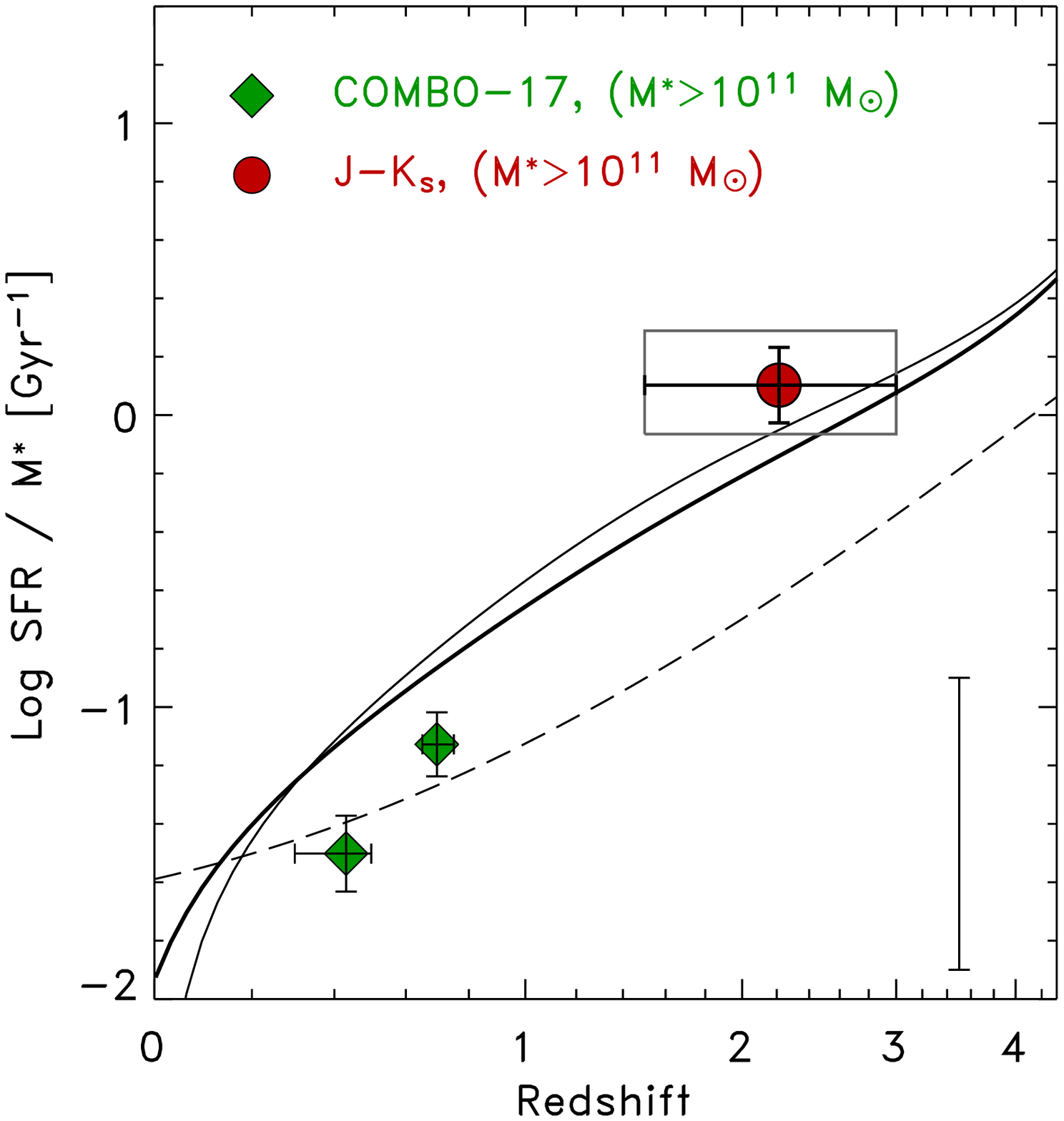}
\end{center}
\caption{Evolution of the integrated specific
SFR, \ie, the ratio of the total SFR to the total stellar mass (from
Papovich et al., 2005).  The curves show the expected evolution of the
ratio of the total SFR to the total galaxy stellar mass densities from
an empirical fit to the evolution of the SFR density \citep[solid
lines, thick line includes correction for dust extinction;][]{col01},
and the model of \citet[dashed line;][]{her03}.   The data points show
results for galaxies with $\mathcal{M}^\ast \geq
10^{11}$~$\mathcal{M}_\odot$. The filled diamonds show the mean values
derived for COMBO--17 galaxies, and the filled circle shows the mean
value for the DRGs.  The box about the DRG point shows how the results
change based on some various assumptions (see text; Papovich et al.\
2005).  The error bars do not include systematic uncertainties in the
SFRs, which are indicated by the inset error
bar. }\label{fig:specsfrevol}\vspace{0pt}
\end{figure*}

Although the sample is biased against galaxies with low stellar masses
(and thus specific SFRs); the data are sensitive to all galaxies with
larger stellar masses.  At $z$$\sim$0.3--0.7, there is an apparent
lack of galaxies with high specific SFRs and high stellar masses.  In
contrast, the massive DRGs at 1.5$\leq$$z$$\leq$3 have much higher
specific SFRs for galaxies.  Quantitatively, the DRGs with
$\mcal$$>$$10^{11}$~\msol\ and 1.5$\leq$$z$$\leq$3 have
$\Psi/\mcal$$\sim$0.2--10 Gyr$^{-1}$, with a mean value of
$\sim$2.4~Gyr$^{-1}$ (excluding X--ray sources).  By $z$$\sim$0.7
(0.4) galaxies with $\mcal$$\geq$$10^{11}$~\msol\ have
$\Psi/\mcal$$\sim$0.1--1 ($\lsim$0.5) Gyr$^{-1}$, an order of
magnitude lower than for the massive DRGs.   We define the integrated
specific SFR as the ratio of the sum of the SFRs, $\Psi_i$, to the sum
of their stellar masses, $\mcal_i$, $\Upsilon$$\equiv$$\sum_i 
\Psi_i / {\sum_i\mcal_i}$,  summed over all $i$ galaxies.  This is
essentially just the ratio of the SFR density to the stellar mass
density for a volume--limited sample of galaxies.
Figure~\ref{fig:specsfrevol} shows the integrated specific SFRs for
DRGs at $z$$\sim$1.5--3.0 and COMBO--17 at $z\sim 0.4$ and 0.7 with
$\mcal \geq 10^{11}$~\msol. The data point for the DRGs includes all
objects with $\mcal$$\geq$$10^{11}$~\msol, assuming that
24~\micron--undetected DRGs have  $f_\nu(24\micron) = 60$~\ujy, and
excluding objects with X--ray detections of IR luminosities of color
indicative of AGN (see Papovich et al.\ 2005).  The upper bound of the
error box shows what happens if we include those objects with possible
AGN.   The bottom bound of the error box shows what happens if we
continue to exclude objects with possible AGN, but assuming that the
24~\micron--undetected DRGs have no star formation.  The integrated
specific SFR in galaxies with \mcal$>$$10^{11}$~\msol\ declines by
more than an order of magnitude from $z$$\sim$1.5--3 to $z$$\lsim$0.7.
This downward evolution in the specific SFRs seems to support the
so--called  ``downsizing'' paradigm.   Our results indicate that
star--formation in massive galaxies is reduced for $z$$\lsim$1 as
galaxies with lower stellar masses have higher specific SFRs.

Figure~\ref{fig:specsfrevol} also shows the specific SFR integrated
over all galaxies (not just the most massive); this is the ratio of
the cosmic SFR density (from Cole et al., 2001) to its integral, i.e.,
$\Upsilon$$=$$\dot{\rho}_\ast / \int \dot{\rho}_\ast\, dt$.   The
global integrated specific SFR declines steadily with decreasing
redshift, \ie, there is a decrease in the global specific SFR. The
evolution in the integrated specific SFR in massive galaxies is
accelerated relative to the global value.   Galaxies with
$\mcal$$\ge$$10^{11}$~\msol\ were forming stars at or slightly above
the rate integrated over all galaxies at $z$$\sim$=1.5--3.   In
contrast, by $z\lsim 1$ galaxies with $\mcal$$\ge$$10^{11}$~\msol\
have an integrated specific SFR much lower than the global value.
Thus, by $z$$\lsim$1.5 massive galaxies have formed most of their
stellar mass, and lower--mass galaxies dominate the cosmic SFR density
(see further discussion in Papovich et al., 2005).

\vspace{-24pt}
\section{Old Stellar Populations in Massive Galaxies at
2$\mathbf{<}$$\mathbf{z}$$\mathbf{<}$3: \hbox{$\;\;\;\;\;\;\;\;\;\;$}
Implications for the IMF at Higher Redshift}
\vspace{-24pt}

Because the DRGs have red colors, the model fits favor
stellar--population ages of 1--2~Gyr. This implies that the earliest
DRG progenitors  formed at $z$$\gsim$5--6, possibly accounting for
much of their stellar mass \citep{for04,pap05}.  Owing mostly to model
degeneracies, the uncertainties on galaxy ages, dust content, and
star--formation histories from the stellar--population modeling can be
more than an order of magnitude \citep[\eg,][]{pap01}.  Nonetheless,
with the model star--formation histories and stellar masses of
galaxies at $z$$\sim$2--3, we can set some broad statistical
constraints on the SFRs of galaxies at $z$$\gsim$5--6.  For example,
\citet{fer02} analyzed the model star--formation histories of
LBGs at $z$$\sim$2--3, and concluded that in order for the progenitors
of these galaxies to produce enough Lyman--continuum photons to
reionize the Universe requires that the stellar populations form from
an IMF heavily weighted toward high--mass stars, and/or an episodic
SFR. \citet{for04} found that if the DRGs had a constant SFR over
their lifetimes, then integrated UV--luminosity density from the DRG
progenitors decreases by a factor of $\approx$2 from $z$$\sim$2 to 6.

\begin{figure*}[t]
\begin{center}\includegraphics*[scale=0.4]{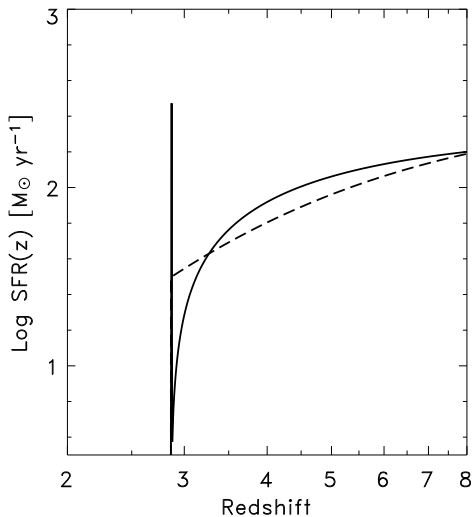}
\end{center}
\caption{Past SFR for the DRG at $z$=2.9 illustrated in
figure~\ref{fig:example_spec}.  The dashed line shows the past SFR for
the model with a monotonically evolving SFR,
$\Psi(t)$$\sim$$\exp(-t/\tau)$.  The solid line shows the past SFR of
a two--component model.  One component is a recent ``burst'', modeled
as a decaying exponential with a short age and e--folding time.  The
other component approximates the average star--formation history of a
``burst'' sometime in the past between $z$=2.9 and
$z$$\sim$$\infty$. }\label{fig:example_pastsfr}\vspace{-6pt}
\end{figure*}

It is interesting to ask what the stellar--population models imply for
the luminosities of the progenitors of the GOODS--S DRGs at
$z$$\gsim$3--5.
%For the DRGs in our sample, we
%can esimate their past SFR using the best--fit models.
Figure~\ref{fig:example_pastsfr} shows the past SFR for the DRG in
figure~\ref{fig:example_spec} for the best-fitting models using the
two different star--formation histories (see
\S~\ref{section:modeling}, figure~\ref{fig:example_spec}).  The dashed
line shows the past UV luminosity for the model with a monotonically
evolving SFR.  The solid line shows the past SFR for the model with
two star formation components.   One component is a monotonically
evolving ``burst'', which in the case of this figure is a very short,
recent burst (causing the spike at $z$$\simeq$3).   The second
component respresents a burst of star formation in the past, averaged
over the time since the Big Bang.   A galaxy with two ``bursts'' of
star--formation would simply show two spikes in
figure~\ref{fig:example_pastsfr}.  If we consider this  model as a
proxy for the history of all galaxies, then the star--formation
history in the figure represents the maximal rate of star formation
due to stochastic bursts as a function of redshift for an individual
galaxy (see Ferguson et al.\ 2002).
 
\begin{figure*}[t]
\begin{center}\includegraphics*[scale=0.4]{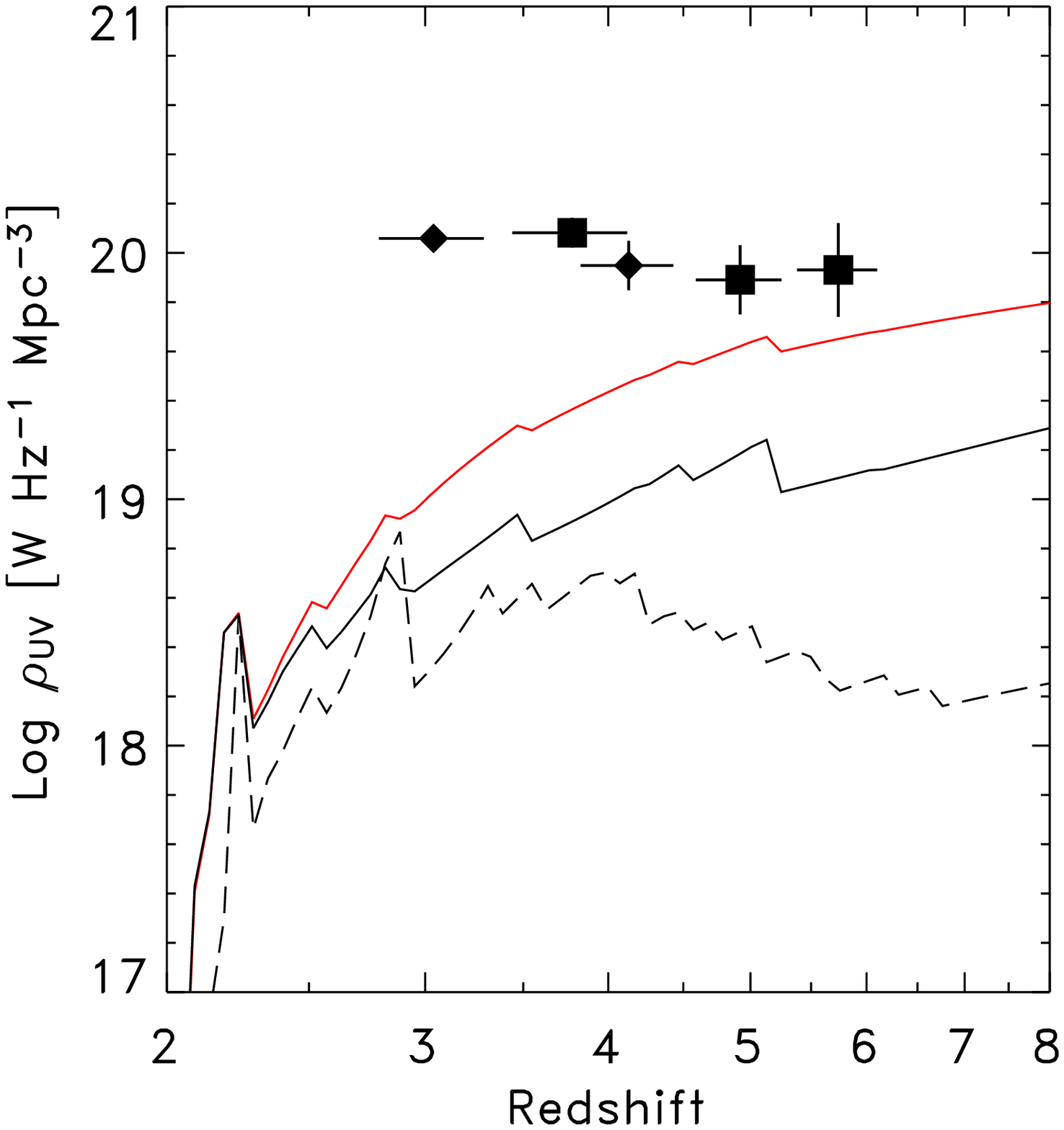}
\end{center}
\caption{Integrated UV luminosity density from the past best--fit
star--formation histories of all DRGs with 2$<$$z$$<$3 and
$\mcal$$\geq$$10^{11}$~\msol.  The dashed line shows the past UV
luminosity density for the single--component star--formation
histories; the solid line shows the two--component models. Both of the
black lines correspond to stellar populations formed with a Salpeter
IMF, with slope $x$=1.35.   The red line indicates the past UV
luminosity density for the two--component models, and an IMF weighted
to high--mass stars, with $x$=0.0.  Data points are from Steidel
\etal\ (1999; diamonds), and Giavalisco et al.\ (2004;
squares).}\label{fig:pastsfr} \vspace{0pt}
\end{figure*}

Figure~\ref{fig:pastsfr} shows the integrated past UV luminosity
density (a proxy for the SFR) for all the GOODS--S DRGs with
2$<$$z$$<$3 and $\mcal \geq 10^{11}$~\msol.  For these redshifts and
stellar mass the sample is approximate complete (i.e., nearly volume
limited) for passively evolving galaxies.  The figure also shows
intantaneous measurements of the UV luminosity density at
$z$$\sim$3--6 from \citet{ste99} and
\citet{gia04b} using LBGs.  Comparing the
measured UV luminosity density to what we infer from the past output
from the DRGs, if the massive DRGs at 2$<$$z$$<$3 formed their stellar
populations with a monotonically evolving SFR, then they account for
$<$5\% of the  $z$$\sim$6 UV luminosity density.  However, if these
galaxies instead form stars in quasi--stochastic bursts throughout
their lifetime, then they account for $\approx$10--15\% of the
$z$$\sim$6 UV luminosity density.  Inverting that statement, roughly
10--15\% of the stars formed in $z$$\sim$6 galaxies end up in
massive DRGs at 2$<$$z$$<$3.

We can go one step further and impose the constraint that the stars
forming at $z$$\sim$6 can not \textit{overproduce} the stellar mass at
$z$$\sim$2--3.  In this case the data allow us to place a very loose
constraint on the IMF of the old stellar populations in the DRGs.  If
the past IMF in these galaxies is weighted to high--mass stars (e.g.,
with a slope shallower than Salpeter, $x$$<$1.35), then one gets more
UV photons produced for a given amount of formed stellar mass (more
``bang for the buck'').   The rate that the light from a stellar
population fades goes as $L_B(t) \sim t^{-\alpha}$.  We approximate
the change in the past UV luminosity for a varying IMF by scaling the
past UV luminosity by the change in $B$--band luminosity.  For a
Salpeter IMF $\alpha = 0.8$, whereas a ``Flat'' IMF (with slope
$x$=0.0) has $\alpha=1.4$.  If the old stellar populations formed with
a top--heavy ($x$=0.0), flat IMF, then the past UV luminosity density is
$\sim$90\% of the $z$$\sim$6
\citet{gia04b} value.  If the measured $z$$\sim$6 UV luminosity
density is in fact lower, then the stellar mass in the DRGs can
account for all \citep{bou05}, or more
\citep{bun04} than what can be formed with a $x$=0.0 IMF.  Therefore,
if the UV luminosity density is lower than the Giavalisco et al.,
(2004b) value, then the IMF \textit{must} have a slope $x$$>$0.0 to
avoid the awkward situation that the $z$$\gsim$6 galaxies overproduce
the stellar mass in DRGs with 2$<$$z$$<$3 and $\mcal$$\geq$$10^{11}$~\msol. 

Such a situation seems extreme.  It is possible that the massive DRGs
contain a large amount of the stars formed at $z$$\sim$6.  Currently,
we have only the limited constraint that the slope of the IMF of the
$z$$\sim$6 star--forming galaxies is $x$$>$0.0.  This may be tightened
as the observations of the $z$$\gsim$6 UV luminosity density improve
(or we decrease the uncertainty in the stellar--population modeling).
Moreover, the constraint will be better if we can measure the UV
luminosity density at yet higher redshift.  For example, Bouwens et al.,
(2004) suggest that the $z$$\sim$7--8 UV luminosity density is lower
than at $z$$\sim$6 by an order of magnitude.  In this case, based on
figure~\ref{fig:pastsfr} we may already be able to discard an IMF with
a very shallow slope.  We may yet achieve indirect constraints on the
IMF of high--redshift galaxies at the epoch of reionization using the
old stellar populations we find at ``low'' redshifts, $z$$\sim$2--3.

\textit{Acknowledgments:}
I wish to thank the many members of the GOODS and MIPS GTO teams for
stimulating conversations, and much hard work on the various aspects
of the program.  I am especially indebted to L.~Moustakas,
M.~Dickinson, E.~Le~Floc'h, G.~Rieke, and E.~Daddi for their help on
this project, and I am thankful for their continued collaboration.  I
also extend my thanks to the conference organizers for
planning a successful meeting; I look forward to the next UCI
symposium.   Lastly, I wish to thank the proceedings editors
(A.~Cooray \& E.~Barton) for their work and considerable patience.
Support for this work was provided by NASA through the \spitzer\ Space
Telescope Fellowship Program, and by NASA through awards issued by JPL.

\end{document}